\documentclass[life,article,accept,moreauthors,pdftex]{mdpi} 
\usepackage{upgreek}
\firstpage{1} 
\pubvolume{1}
\issuenum{1}
\articlenumber{0}
\pubyear{2021}
\copyrightyear{2020}
\datereceived{} 
\dateaccepted{} 
\datepublished{} 
\hreflink{https://dx.doi.org/10.3390/life101000} % Please use \linebreak if need to do line break
\Title{Super-Earths, M Dwarfs, and Photosynthetic Organisms: Habitability in the Lab}%Attention AE/ME. The following layout issues have not been checked by the English Editing Department and must be carefully verified by the AE/Layout Department: All callout issues, bold usage of callouts, and references to callouts in the text. Correct callout usage in figures. Figure and Table layout issues. Footnote formatting and Glossaries have not been checked. En dash usage for negative values, en dash usage to indicate relationships, en dash usage to indicate bonds (especially in chemistry). The English Editing Department is not responsible for correct italic usage for genes, proteins and technical terminology. This responsibility belongs to the authors. The following are also not checked: spacing between numbers and units of measurement, ratios, en dashes for ranges, date and time formats, punctuation in equation lines, and less than/more than spacing (< >). Finally, capitalization and layout of titles/headings must be properly checked as well as ensuring 'Eq.' and 'Fig.' are properly spelled out, as these are layout issues.

\TitleCitation{Super-Earths, M Dwarfs and Photosynthetic Organisms: Habitability in the Lab}
\newcommand{\nat}{Nature}

\newcommand{\apj}{Astrophys. J.}
\newcommand{\apjl}{Astrophys. J. Lett.}
\newcommand{\apjs}{Astrophys. J. Suppl.}
\newcommand{\araa}{Annu. Rev. Astron. Astrophys.} 
\newcommand{\aap}{Astron. Astrophys.}
\newcommand{\icarus}{Icarus}
\newcommand{\mnras}{Mon. Not. R. Astron. Soc.}

\newcommand{\physrep}{Phys. Rep.}

\newcommand{\grl}{Geophys. Res. Lett.}

\Author{Riccardo Claudi%Please carefully check the accuracy of names and affiliations. RCl: CONFIRMED
 $^{1,}$*\orcidA{}, Eleonora Alei $^{1,2}$\orcidB{}, Mariano Battistuzzi $^{3,4}$\orcidC{}, Lorenzo Cocola $^{5}$\orcidD{}, Marco Sergio Erculiani %MDPI: please add complete author name for E. Alei.
 $^{6}$, Anna Caterina Pozzer $^{1, 3}$\orcidE{}, Bernardo Salasnich $^{1}$\orcidF{}, Diana Simionato $^{3}$, Vito Squicciarini $^{1,7}$, Luca Poletto $^{5}$\orcidG{} and Nicoletta La Rocca $^{3,4}$\orcidH{}}
\AuthorNames{R. Claudi, E. Alei, M. Battistuzzi, L. Cocola, M. S. Erculiani, A. C. Pozzer, B. Salasnich, D. Simionato, V. Squicciarini, L. Poletto and N. La Rocca}
\AuthorCitation{Claudi, R.; Alei, E.; Battistuzzi, M.; Cocola, L.; Erculiani, M.S.; Pozzer, A.C.; Salasnich, B.; Simionato, D.; Squicciarini, V.; Poletto, L.; Rocca, N.L.}
\address{%
$^{1}$ \quad Osservatorio Astronomico di Padova, INAF, 35122 Padova, Italy; %Please confirm the newly added City, postal code and Country Name. RCl: CONFIRMED
 Elalei@phys.ethz.ch (E.A.); caterina.pozzer@inaf.it (A.C.P.); bernardo.salasnich@inaf.it (B.S.); vito.squicciarini@inaf.it (V.S.)\\
$^{2}$ \quad Institute for Particle Physics and Astrophysics, ETH Zurich, 8093 Zurich, %Please confirm the newly added City, postal code. RCl: CONFIRMED
 Switzerland\\
$^{3}$ \quad Department of Biology, University of Padova, 35131 Padova, Italy; mariano.battistuzzi@phd.unipd.it (M.B.); diana.simionato@unipd.it (D.S.); nicoletta.larocca@unipd.it (N.L.R.)\\%Please confirm the newly added postal code. RCL: Modified the right one is 35131
$^{4}$ \quad Centro di Ateneo di Studi e Attivit\`a Spaziali (CISAS) \textit{Giuseppe Colombo}, 35131 Padova, Italy\\ 
$^{5}$ \quad Institute for Photonics and Nanotechnologies, CNR, 35131 Padova, Italy; lorenzo.cocola@cnr.it (L.C.); luca.poletto@cnr.it (L.P.)\\ 
$^{6}$ \quad Istituto di Radioastronomia, INAF, 40129 Bologna, Italy; marco.erculiani@inaf.it\\
$^{7}$ \quad Department of Physics and Astronomy, University of Padova, 35121 Padova, Italy\\}

\corres{Correspondence: riccardo.claudi@inaf.it}

\abstract{In a few years, space telescopes will investigate our Galaxy to detect evidence of life, mainly by observing rocky planets. In the last decade, the observation of exoplanet atmospheres and the theoretical works on biosignature gasses have experienced a considerable acceleration. The~most attractive feature of the realm of exoplanets is that 40\% of M dwarfs host super-Earths with a minimum mass between 1 and 30 Earth masses, orbital periods shorter than 50 days, and radii between those of the Earth and Neptune (1--3.8 R$_\oplus$). Moreover, the recent finding of cyanobacteria able to use far-red (FR) light for oxygenic photosynthesis due to the synthesis of chlorophylls $d$ and $f$, extending in vivo light absorption up to 750\ nm, suggests the possibility of exotic photosynthesis in planets around M dwarfs. 
Using innovative laboratory instrumentation, we exposed different cyanobacteria to an M dwarf star simulated irradiation, comparing their responses to those under solar and FR simulated lights.~As expected, in FR light, only the cyanobacteria able to synthesize chlorophyll $d$ and $f$ could grow. Surprisingly, all strains, both able or unable to use FR light, grew and photosynthesized under the M dwarf generated spectrum in a similar way to the solar light and much more efficiently than under the FR one. Our findings highlight the importance of simulating both the visible and FR light components of an M dwarf spectrum to correctly evaluate the photosynthetic performances of oxygenic organisms exposed under such an exotic light~condition.}


\begin{document}
%%%%%%%%%%%%%%%%%%%%%%%%%%%%%%%%%%%%%%%%%%

%%%%%%%%%%%%%%%%%%%%%%%%%%%%%%%%%%%%%%%%%%
%\setcounter{section}{-1} %% Remove this when starting to work on the template.
\section{Introduction}
\label{sec:introduction}

The considerable number of new worlds discovered so far is pushing scientists to provide evidence of life on other planets. The diversity in kind, composition, masses, and~radii of these new worlds is so vast that almost all possible mass values are covered in continuity from Mars ($0.11\ \text{M}_\oplus$) up to super-Jupiters (>10 $\text{M}_\text{J}$). Among all the planetary hosts, low mass stars, mainly M spectral type stars, are the main targets of the extrasolar planet surveys due to both their high density in the Galaxy and their small radii that provide higher amplitude transit signals than solar-like stars \cite{dressingandcharbonneau2013apj, davenportetal2016apj}.~Indeed, the most attractive characteristic of these systems is that 40\% of M stars host super-Earths with a minimum mass between about 1 and 30 Earth masses, orbital periods shorter than 50~days, and~radii between those of the Earth and Neptune (1--3.8 R$_\oplus$). 
Due to these high occurrence rates, super-Earths (1--10 M$\oplus$) represent the most common type of components of planetary systems in the Galaxy \cite{mayoretal2014nature}. Even more striking, the frequency of super-Earths found in the habitable zone (HZ) of M dwarfs (with a period between 10 and 100 days) is about \mbox{50\% \cite{bonfilsetal2013aa, kopparapuetal2013apj}.}
These results renew, with higher and more interdisciplinary efforts, the~search for life as an astrophysical problem. 
In this framework, it is critical to determine the types of biosignatures (based on the so-far-recognized life signatures) for when designing the next generation of ground- and space-based instruments that will observe these planets at both high spectral and spatial resolutions (e.g., Reference~\cite{seageretal2012asbio, seageretal2016asbio, kaltenegger2017araa, grenfell2017phr, catlingetal2018asbio, schwietermanetal2018asbio, claudiandalei2019galax, lingamandloeb2019rvmp}).~Among all the biosignatures, oxygen~seems to be the most prominent signature that can reveal the existence of life on other planets \cite{lovelock1965natur}. In this sense, it is our most robust and the most studied biosignature \cite{meadowsetal2018asbio}.
Its presence, together with other gases, like CH$_4$ or N$_2$O, is the signal of a thermodynamic disequilibrium that, for a long time, has been considered as compelling evidence for life (e.g., Reference~\cite{lederberg1965natur}). 
Since the closure of of Seager \& Bains (2015) \cite{seagerandbain2015scia}, many arguments countered this concept because using the thermodynamic disequilibrium as a biosignature cannot be easily generalized. Other authors have also argued that oxygen is not a suitable bioindicator due to photochemical reactions that have abiotic O$_2$ as a byproduct (e.g., H$_2$O and CO$_2$ photoionization). A detailed discussion of several possible false positives is presented in Harman et al. (2015) \cite{harmanetal2015apj}. 

Today, in the the atmosphere of Earth, oxygen is highly abundant (21\% by volume) due to the oxygenic photosynthesis of plants, algae, and cyanobacteria, the presence of which can be detected by remote observations, not only for the presence of O$_2$ in the atmosphere of the planet but also by discerning the red edge. The red edge is a feature associated with the high reflectance of photosynthetic organisms at near infra-red (NIR) in contrast with the absorption by chlorophyll in wavelengths shorter than about 700 nm. This phenomenon emerges due to the scattering of light at the interfaces between the cell walls and the air space inside the organism \cite{gatesetal1965apopt}. 

In recent years, biologists have found species of cyanobacteria able to use far-red (FR) light for oxygenic photosynthesis due to the synthesis of chlorophylls $d$ and $f$ \cite{ganandbryant2015envmi}, extending in~vivo light absorption up to 750\ nm, suggesting the possibility of exotic photosynthesis in planets around M stars. So far, a number of works have discussed the possibility of emerging oxygenic photosynthesis on a planet in the HZ of an M star (e.g., Reference~\cite{wolstencroftandraven2002icar, tinettietal2006apj, kiangetal2007asbio}) under favorable conditions. 

Considering all these favorable observational and theoretical circumstances, it is important to assess, in an experimental way, the consequences of oxygenic photosynthesis on planets orbiting in the HZ of M stars. 
In this paper, we present the laboratory set up and experiments conducted with the aim of understanding the performance of photosynthetic organisms exposed to conditions similar to that of an Earth-like planet in the HZ of an M~star. In particular, the laboratory set up simulates the exoplanetary surface temperature and radiations. In these experiments, we analyzed the growth and the photosynthetic efficiency of several cyanobacteria with different photosynthetic behaviors, based on chlorophyll fluorescence measurements. In particular, we selected a cyanobacterium unable to utilize FR light for carrying out oxygenic photosynthesis, two species able to exploit it and a species both able to use FR light and perform the so-called Chromatic Acclimation (CA), changing its color depending on the incident light spectrum. 
%We also made preliminary experiments to study how the growth of cyanobacteria and the O$_2$ - CO$_2$ production - consumption differ under different light regimes.

This paper is organized as follows: In Section\ \ref{sec:experiment}, we describe the approach to the problem and the experimental plan. 
%the science case and the astrophysical and astrobiological framework
In Section\ \ref{sec:labsetup}, the experimental set up and its validation is described; in Section\ \ref{sec:conc}, we discuss the microorganisms we used and report the results of the experiment. Section \ \ref{sec:disconcl} is allocated to the discussion and the conclusion.
%%%%%%%%%%%%%%%%%%%%%%%%%%%%%%%%%%%%%%%%%%
\section{Experimental Aims}
\label{sec:experiment}
Our research plan aim was analyzing the changes caused by the presence of photosynthetic bacteria to the chemical composition of a simulated, secondary atmosphere of a super-Earth inside the HZ of an M star. This expression of intents delimits the environmental parameters (pressure, temperature, irradiation pattern, initial atmospheric chemical composition) we want to simulate in the laboratory. The description of another version of this experiment is detailed in Reference \cite{claudietal2016Ijasb}. Here, we describe an entirely new evolution of the experiment, putting it into context.
%A complete description of the main astrophysical and astrobiological framework of the experiment is given in Reference \cite{claudietal2016Ijasb} together with detailed planning of the several time-steps of the experimental work. Here, we give a summary in order to put the experiment into the context. 

\subsection{The Astrophysical Context}
\label{sec: apcont}
\begin{figure}[H]
\includegraphics[width=13 cm]{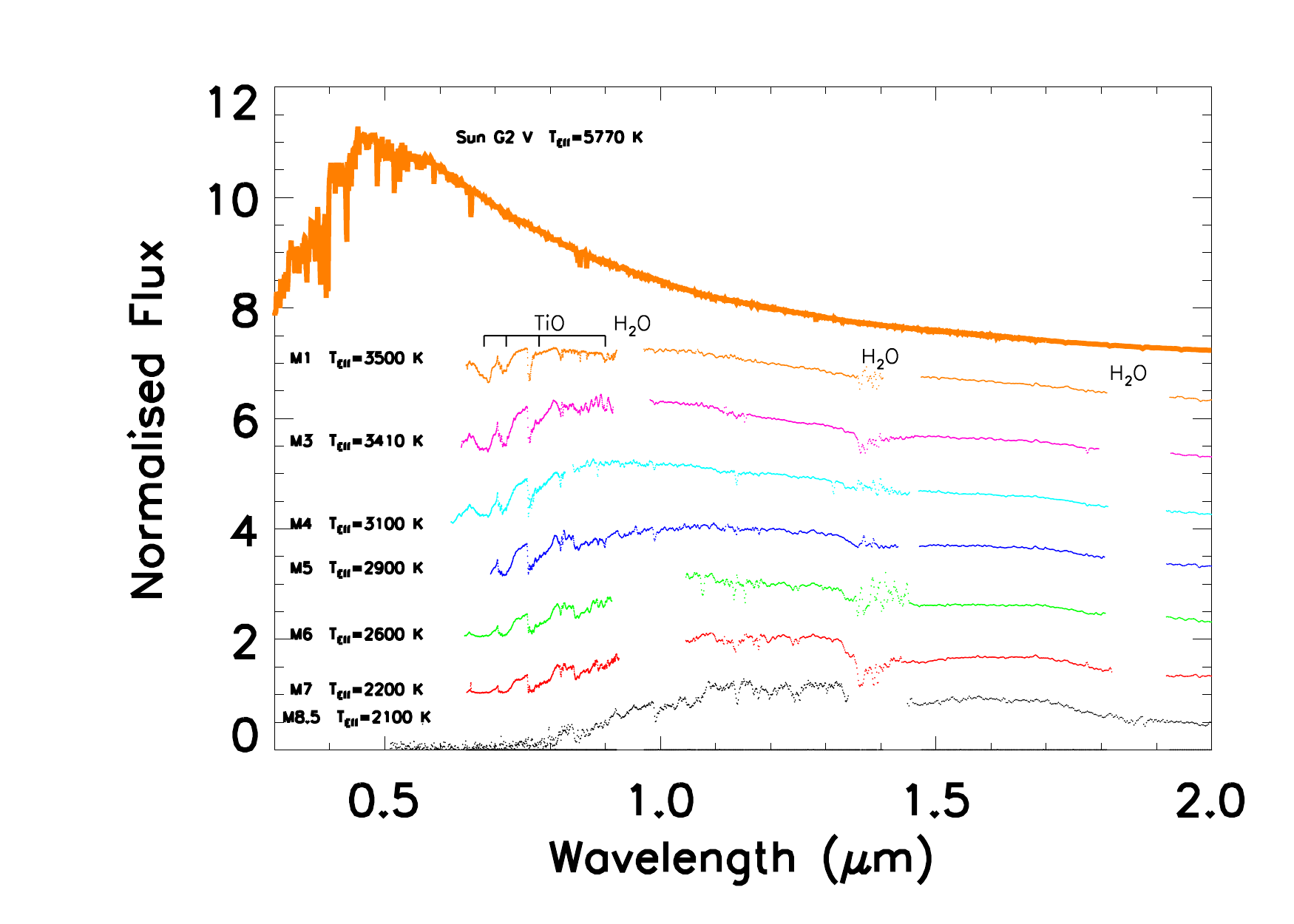}
\caption{Examples of various M-stars spectra \cite{leggettetal2000apj}, from higher to lower photospheric tempretures. The main molecular bands depleting the visible and near infra-red (NIR) flux are also indicated. The spectral features of M-stars are indicated in Reference \cite{leggettetal1996apjs}.~The upper part of the figure shows the spectrum of the Sun for comparison. The Sun and M star spectra have been normalized to the flux they emit at $1.2\ \upmu$m and~offset.}
\label{fig:mstarspectra}
\end{figure}

M dwarfs are faint hydrogen-burning stars with masses ranging between $0.6$--$0.075\ \text{M}_\odot$ 
and very low photospheric temperatures (3870\ K for an M0 V and 2320\ K for an M9 V star \cite{pecautetal2013apjs}). Their spectra are characterized by maximum emission at longer wavelengths than those of the Sun and molecular band absorptions that deplete the emitted visible flux (see Figure\ \ref{fig:mstarspectra}). 
In the visible, the Sun (G2 V) has an absolute magnitude of about 4.8, while the M stars are about 4 orders of magnitude fainter than the sun (the absolute visual magnitude of an M star ranges between 9 for M0 V to 20 mag for M9 V \cite{pecautetal2013apjs}).
%In the visible, these stars are about 4 orders of magnitude fainter than the Sun (the absolute visual magnitude of an M star ranges between 9 to 20 mag). 

In other words, M dwarfs are very different from the Sun in both luminosity and spectral distribution. Luminosity affects the position of the HZ around the star, whereas~spectral distribution influences the number of photons available in the PAR (Photochemically Active Radiation) wavelength range ($400\leq \ \lambda \leq \ 700$\ nm), challenging the possibility of photosynthesis on the surface of a rocky planet orbiting in the HZ of this kind of star. 

Not only the wavelength range of the spectral peak, but also the qualitative shape of the spectrum and how it interacts with an HZ planetary atmosphere, vary significantly across the M star mass range. In fact, due to the faintness of M stars, the HZ is as close as $\sim$0.30 au or even closer to the star \cite{kastingetal1993icar, selsisetal2007aa, kopparapuetal2013apj}. Hence, the planet orbiting inside the HZ results as tidally locked and becomes a synchronous rotator with 1:1 spin-orbit resonance (e.g., Reference~\cite{kastingetal1993icar}) or higher (3:2) (e.g., Reference~\cite{brownetal2014ijasb}).
Another possible consequence to consider is whether there is sufficient flux in the wavelength range used for photosynthesis \cite{lingamandloeb2019mnras}. Heath et al. (1999)~\cite{heathetal1999oleb} determined that the PAR from M stars would be lower than the average terrestrial value by about one order of magnitude. 

In past years, several authors have also argued that terrestrial planets within the HZs of M dwarfs may not be habitable. The main reasons range from the possible deficiency of volatiles in the planetary atmosphere \cite{lissauer2007apjl}, to the scarce water delivery during the planet evolution \cite{raymondetal2007apj}, or the loss of planetary water during the pre-main sequence due to the higher luminosity of the protostar in that evolutionary phase \cite{lugerandbarnes2015asbio}. The latter seems a showstopper indeed. In any case, Tian and Ida (2015) \cite{tianandida2015natge} showed that the content of water in Earth-like planets orbiting low mass stars could be rare, but dune and deep ocean worlds may be common.

Nevertheless, many authors are optimistic asserters that the oxygenic photosynthesis can take place on a super-Earth surface also under these conditions. Recent studies on possible water loss in the atmosphere of planets orbiting very cool stars, like Trappist-1 d, show that these planets may still have retained enough water to support surface habitability \cite{gillonetal2016nature, bolmonetal2016aa}. Furthermore, previous work on the photosynthetic mechanisms and spectral energy requirements elucidated that photosynthesis can still occur in harsh and photon-limited environments (e.g., Reference~\cite{heathetal1999oleb,wolstencroftandraven2002icar, tinettietal2006apj, kiangetal2007asbio, scaloetal2007asbio}).

Another possible showstopper discussed in several works is the obstacle caused by the stellar activity. 
M dwarfs, by nature, are characterized by their high stellar activity. These stars can significantly change their activity depending on their evolutionary stage. During a quarter of their early life, M dwarfs release high amounts of XUV %Please define if appropriate.
through flares and chromospheric activity \cite{kiangetal2007asbio}, while quiescent stars emit little UV radiation and have no flares~\cite{seguraetal2005asbio}. Planets orbiting around M dwarfs often receive high doses of XUV radiation during stellar flares, with fluxes that can increase ten to a hundred times and occur \mbox{10--15~times} per day. These events rapidly change the radiation environment on the surface~\cite{seguraetal2010asbio, franceetal2013apj, cuntzandguinam2016apj} and possibly erode the ozone shield, as well as parts of the atmosphere. However, some researchers point out that these planets could remain habitable despite these issues \cite{seguraetal2010asbio, omalleyjamesandkaltenegger2017lpico, omalleyjamesandkaltenegger2019mnras}. The presence of a strong magnetic field or a thick \mbox{atmosphere \cite{tarteretal2007asbio, galeandwandel2017ijasb}} could avoid planetary atmosphere erosion. On these planets, possible organisms could develop UV protecting pigments and DNA repair mechanisms, similar to Earth, or thrive in subsurface niches \cite{ranjanandsasselov2016asbio} and underwater, where radiation is less intense.~This kind of life, of course, would not be detectable remotely. A UV-protective mechanism that could be detected remotely is the photo-protective bioluminescence, where proteins shield the organisms from dangerous UV radiations, emitting the energy at longer, detectable wavelengths. Moreover, it would allow organisms to live on the surface of planets orbiting active stars and is a mechanism already at play on Earth for some species of corals \cite{ omalleyjamesandkaltenegger2019mnras}. {The flaring activity of M dwarfs, in particular, could even be beneficial for oxygenic photosynthesis, if the XUV portion is adequately shielded by the atmosphere. They could indeed enhance the effectiveness of oxygenic photosynthesis due to the additional flux in the PAR that they can generate \cite{mullanandbais2018apj}, even if they are not thought to allow an Earth-like biosphere in planets orbiting the HZ of the majority of the M dwarfs known \cite{lingamandloeb2019rvmp}}.

From an observational point of view, \textit{Kepler} found that Earth-sized planets (1.0--1.5~$\text{R}_\oplus$) are common around M stars with an occurrence rate of 56\% with periods shorter than 50\ days. Super-Earths with radii between 1.5--2.0 $\text{R}_\oplus$ and periods shorter than 50\ days orbit M dwarfs with an occurrence rate of 46\% \cite{dressingandcharbonneau2015apj}. Similar high occurrence rates are reported by the radial velocity surveys \cite{mayoretal2014nature}. Notable examples are the Trappist-1 system with seven super-Earths \cite{gillonetal2017nature} and Proxima Cen\ b \cite{angladaescudeetal2016nature} orbiting two M stars.

These planets are predicted to have large surface gravities ($25\ \text{ms}^{-2}$ for 5 M$_\oplus$ (Reference \cite{millerriccietal2009apj} and References therein)) and are likely to exist within a wide range of atmospheres: some of them could be able to retain a thick H-rich atmosphere, whereas others could have a stronger resemblance to Earth with heavier molecules in their atmospheres.~A third possibility could be an atmosphere with a moderate abundance of hydrogen due to its escape and/or molecular hydrogen outgassing \cite{millerriccietal2009apj}. 

\subsection{The Experiment Plan}
\label{sec:exptim}
The background scenario is crowded with theoretical hypotheses on the photosynthesis at work on planets orbiting M stars; however, to the knowledge of authors, no~experimental work has been performed directly exposing photosynthetic organisms to a simulated M-dwarf spectum.~Here, the (unavoidable) working hypothesis is that the evolution of extra-terrestrial life converged to pigment production and photosynthetic mechanisms similar to that of terrestrial extremophiles under non-Earth conditions. 

Our simple approach consists of the following steps:
\begin{enumerate}[leftmargin=22pt,labelsep=2pt]
\item[i] We set up the laboratory instrumentation and selected the organisms for the tests (see~Section\ \ref{sec:orga}).~We built some of the laboratory tools ex-novo. We first built a light source suitable for the purpose of the experiment, simulating the star irradiation (see~Section\ \ref{sec:illumina}). Secondly, we built the reaction cell (see Section\ \ref{sec:reactcell}).
\item[ii] Before conducting the main experiment, we performed a fiducial one considering the terrestrial environment. We irradiated the selected organisms with solar light and within a terrestrial atmosphere environment.
\item[iii] \textls[-5]{Once we checked that the experimental set up functions well in terrestrial condition, we~switched to the M star irradiation of organisms, considering a \mbox{terrestrial~atmosphere.}}
\item[iv] Lastly, we are planning to conduct experiments using the M star light to irradiate the cyanobacteria that will be put in a modified atmosphere.~The composition is defined using the 1-D model of the atmosphere of super-Earths described by \mbox{Petralia et al. 2020 \cite{petraliaetal2020mnras}} and Alei et al. 2020 (submitted). 
\end{enumerate}

In this paper, we address the first three points. The fourth one will be part of future~work.
%%%%%%%%%%%%%%%%%%%%%%%%%%%%%%%%%%%%%%%%%%

\section{Laboratory Set Up}
\label{sec:labsetup}

\begin{figure}[H]
\includegraphics[width=12 cm]{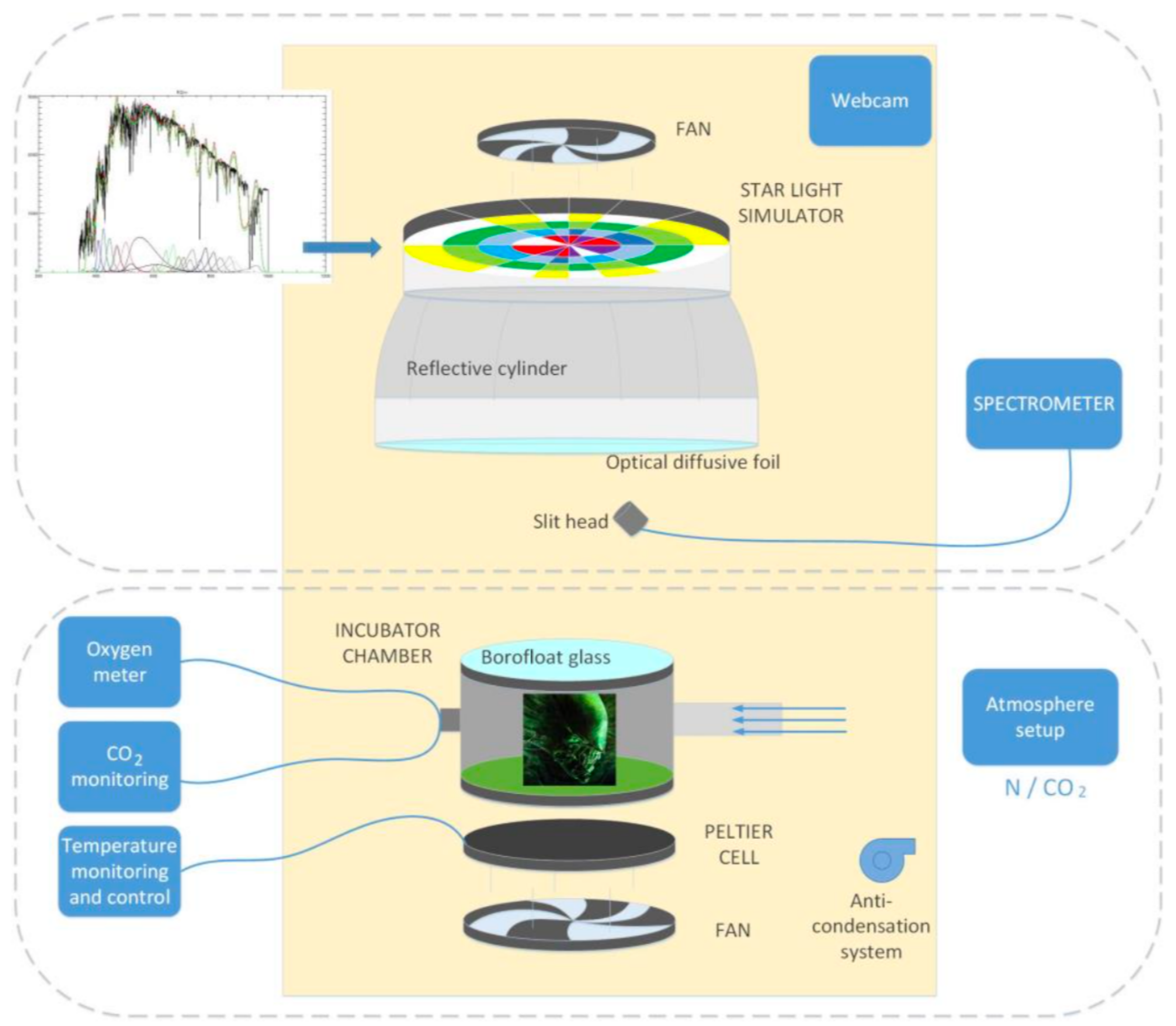}
\caption{The instrumental set up. The upper part shows the starlight simulator (composed of the illuminator and a spectrometer); the bottom panel shows the incubator chamber, where the photosynthetic microorganisms used for the experiments are hosted.}
\label{fig:setupsketch}
\end{figure}

The experimental set up is sketched in Figure\ \ref{fig:setupsketch}. It can be conceptually split into two parts: the stellar simulator on top, composed of the illuminator and a spectrometer, with~the reaction cell on the bottom, where the microorganisms used for the experiments are hosted. The entire system is isolated in a dark container and is cooled by two fans, it is equipped with an anti-condensation system, and it is monitored through a webcam.~A~control PC %Please define if appropriate.
allows to operate remotely the system without interfering with the experiment.

\subsection{Star Irradiation Simulator}
\label{sec:illumina}
To achieve the described experimental aim, it was necessary to have an unconventional light source. In particular, the light sources used in photosynthetic study facilities are mainly metal halide, high-pressure sodium, fluorescent, and incandescent lamps. These~lamps are commercially available and mostly emit solar or close to solar radiation, with a limited capability in adjusting the color temperature and the intensity of the output radiation.~In our experiment, we need a light source able to reproduce the irradiation of stars other than the Sun in a quite simple and direct way. Furthermore, it should be operable without interfering with the experiment. To achieve our goal, we designed a completely different light source using light-emitting diodes (LEDs) controlled remotely by a computer. 

LEDs have also been used in the laboratory as light sources for their efficiency in plants growth \citep{bulaetal1991hortscience}. For that purpose, the LED-based devices are built to illuminate the material in the red part (600--700 nm) of the PAR, while, typically, the blue part of PAR is covered by blue fluorescent lamps.

For our purposes, we need a completely different device. In fact, for Wien's law ($\lambda_{\text{Max}}T=290 \times 10^{4}$\ nm K), different stars of different spectral types have the maximum of their emission at different wavelengths. In particular, while the Sun has a peak of emission at about 550 nm, an M star, which is about a factor of 2 cooler than the Sun, has its emission peak in the NIR range (at about 1000\ nm). In order to appreciate the differences in the slope of the spectra of the stars of different spectral types, we need a collection of LEDs able to cover a slightly longer wavelength range than the PAR, between about 350--1000~nm. Moreover, this device shall be able to modulate the LEDs intensity in order to mimic, as~close as possible, the flux variation of stars of different spectral types. 

The available LEDs allow us to consider the wavelength range between 365\ nm and 940\ nm covered by 25 dimmable channels (see Table\ \ref{tab:leds}). Because LEDs covering such wavelength range are manufactured with different technologies (from AlGaN%Please define if appropriate.
/InGaN %Please define if appropriate.
to GaAs%Please define if appropriate.
/InGaP%Please define if appropriate. RCl: they are all name of molecules
), their emitted luminosities are also different from each other, and each channel has a different number of LEDs to achieve the required optical power at a specific wavelength. Furthermore, we added a white LED with a correlated color temperature (CCT) of 2200--2780 K to fill the spectrum in the 630\ nm region. We used 312 LEDs in total, arranged in five concentric rings on which the mosaic of circuit boards is arranged in a pie-chart shape, on the surface of which the diodes have been welded \cite{trivellinetal2016spie}. Each~channel is tunable enough to allow us to reproduce the radiation of stars of F, G, K, and M spectral~types.

The modularity design of the board permits easy maintenance in case of damage, allowing us to remove only the problematic piece.~The disposition of diodes on the board was designed to reduce the non-uniformity of the flux, due to the intrinsic light exit angle of each led. Moreover, a reflective cylinder and an optical diffusive foil were mounted to increase the uniformity. Since the thermal power of the system dominates its radiation power, the diodes are cooled by a fan set on the back of the board.
A spectrometer collects the light through a slit head placed at a manually adjustable distance from the diffusive foil. The adopted spectrometer is a Component Off The Shelf (COTS) component. We selected the FLAME %Please define if appropriate.
VIS%Please define if appropriate.
-NIR by Ocean Optics, in which its $2040 \times 2040$ pixel detector covers the wavelength range 190--1100~nm. 

% start a new page without indent 4.6cm
\clearpage
\end{paracol}
\begin{specialtable}[H]
\widetable
\caption{The twenty-five channels of the star irradiation simulator, together with the measured $\lambda_{peak} $ and the light-emitting diode (LED) codes.}
\label{tab:leds}
\scalebox{0.82}[0.82]{\begin{tabular}{ccccl}
\toprule
\textbf{Nominal	} \boldmath{$\lambda_{peak}$} \textbf{(nm)}  & \textbf{Measured} \boldmath{$\lambda_{peak} \pm 4\%$} \textbf{(nm)}                    & \textbf{LEDs  Number    } &  \textbf{Tot. Luminosity @0.7A\&25}\boldmath{$^\circ$} \textbf{(W)}	&  \textbf{LED Type}\\
\midrule
365.0			&	368.0 			 &     5  	&	2.735		& Engin LZI-00U600\\   
385.0			&	390.0 			&    15 	&	2.140		& Lumileds LHUV-0380-0200 \\
405.0			&	404.0   			&     10	&	6.030		& Lumileds LHUV-0400-0500\\
425.0			&	424.0			&	10	&	8.770		& Lumileds LHUV-0420-0650 \\
447.5			&	450.0			&	10	&	6.980		& Lumileds LXZ1-PR01       \\
470.0			&	476.0			&	10	&	4.670		& Lumileds LXZ1-PB01       \\
485.0			&	469.0			&	7	&	5.470		& Osram CRBP-HXIX-47-1 \\
505.0			&	499.0			&	25	&	7.400		& Lumileds LXZ1-PE01	\\
530.0			&	520.0			&	10	&	2.260		& Lumileds LXZ1-PM01	\\
567.5			&	548.0			&	45	&	22.500		& Lumileds LXZ1-PX01	\\
590.0			&	604.0			&	20	&	1.530		&  Lumileds LXZ1-PL01	\\
627.0			&	634.0			&	10	&	3.780		&  Lumileds LXZ1-PD01	\\
655.0			&	665.0			&	10	&	4.150		&  Lumileds LXZ1-PA01	\\
680.0			&	689.0			&	14	&	2.814		&  Roithner SMB1N-680	\\
700.0			&	708.0			&	10	&	2.070		&  Roithner SMB1N-700	\\
720.0			&	727.0			&	11	&	2.220		&  Roithner SMBIN-720D	\\
740.0			&	738.0			&	8	&	4.100		&  Engin LZ1-00R300	\\
760.0			&	763.0			&	6	&	2.390		& Roithner SMB1N-760D	\\
780.0			&	777.0			&	8	&	3.390		& Roithner SMB1N-780D	\\
810.0			&	807.0			&	8	&	4.060		&  Roithner SMB1N-810D	\\
830.0			&	834.0			&	15	&	3.860		&   Roithner SMB1N-830N\\
870.0			&	871.0			&	6	&	3.920		& Osram SFH 47155 \\
880.0			&	889.0			&	16	&	4.700		&  Roithner SMB1N-880\\
940.0			&	972.0			&	9	&	5.820		& Osram SFH 4725S \\
white (2200 K)		&	605.0			&	14	&	5.820		& Lumileds 997-LXZ1-2280-5-2200 \\
\bottomrule
\end{tabular}}

\end{specialtable}%
\begin{paracol}{2}
%\linenumbers
\switchcolumn

The illuminator is controlled by a custom control software \cite{salasnichetal2018spie} that, by means of a graphical user interface (GUI), allows the user to select an appropriate spectrum chosen from a spectral library. 
For the input spectrum, the control software calculates the intensities of the 25 channels to best fit the spectrum. In any case, through the GUI, the user has access to each channel of the illuminator setting the output flux of the channel.~The set spectrum is shown in a window of the GUI. The emitted spectrum is registered by the spectrograph and is superimposed on the input spectrum. Slight differences between the two can be fixed by adjusting the luminosity of each channel. The left panel of Figure\ \ref{fig:guillu} shows the simulated spectrum of a solar star (light SOL%Please define if appropriate.
; see Section\ \ref{sec:grow}), while the right panel of the same figure shows the simulated M star spectrum (light M7; see Section\ \ref{sec:grow}). In both panels the input spectrum is represented in red color, and the emitted spectrum in blue. The input spectra are smoothed (e.g., see Figure\ S1 in the Supplementary Materials for an M7 V star) due to the difficulties in reproducing the high resolution stellar spectra by the spectrum~simulator.

\subsection{The Reaction Cell}
\label{sec:reactcell}
The incubator cell (see Figure\ \ref{fig:recell}) is a steel cylinder of 0.5\ l of volume in which the light enters through a thermally resistant Borofloat glass, with over 90\% transmission in 365--940 nm wavelength range. 
The atmosphere in the cell can be flushed to change the initial O$_2$, CO$_2$, and N$_2$ levels. The cell is provided with pipe fittings and connected to an array of flow meters and needle valves (each for a different input gas: N$_2$, O$_2$, CO$_2$) to inject atmospheres of controlled and arbitrary compositions. Once the desired mixture is flushed through the cell, the input and output valves are closed to seal the inside environment and leave it to its evolution. Water vapor will quickly reach saturation value, due to the water-based medium in the sample Petri dish. When using high carbon dioxide levels, caution should be exercised to ensure a long enough flushing time to achieve equilibrium between the gas phase and dissolved CO$_2$ due to its high solubility. The base of the cell and the sample are kept at a constant temperature by controlling a Peltier cell on which the cylinder is leaned. The Peltier temperature set point is always kept $2^\circ$ C lower than the surrounding environment (which its temperature is also controlled) to avoid any condensation on the upper glass window. 

\begin{figure}[H]
\includegraphics[width=13.5 cm]{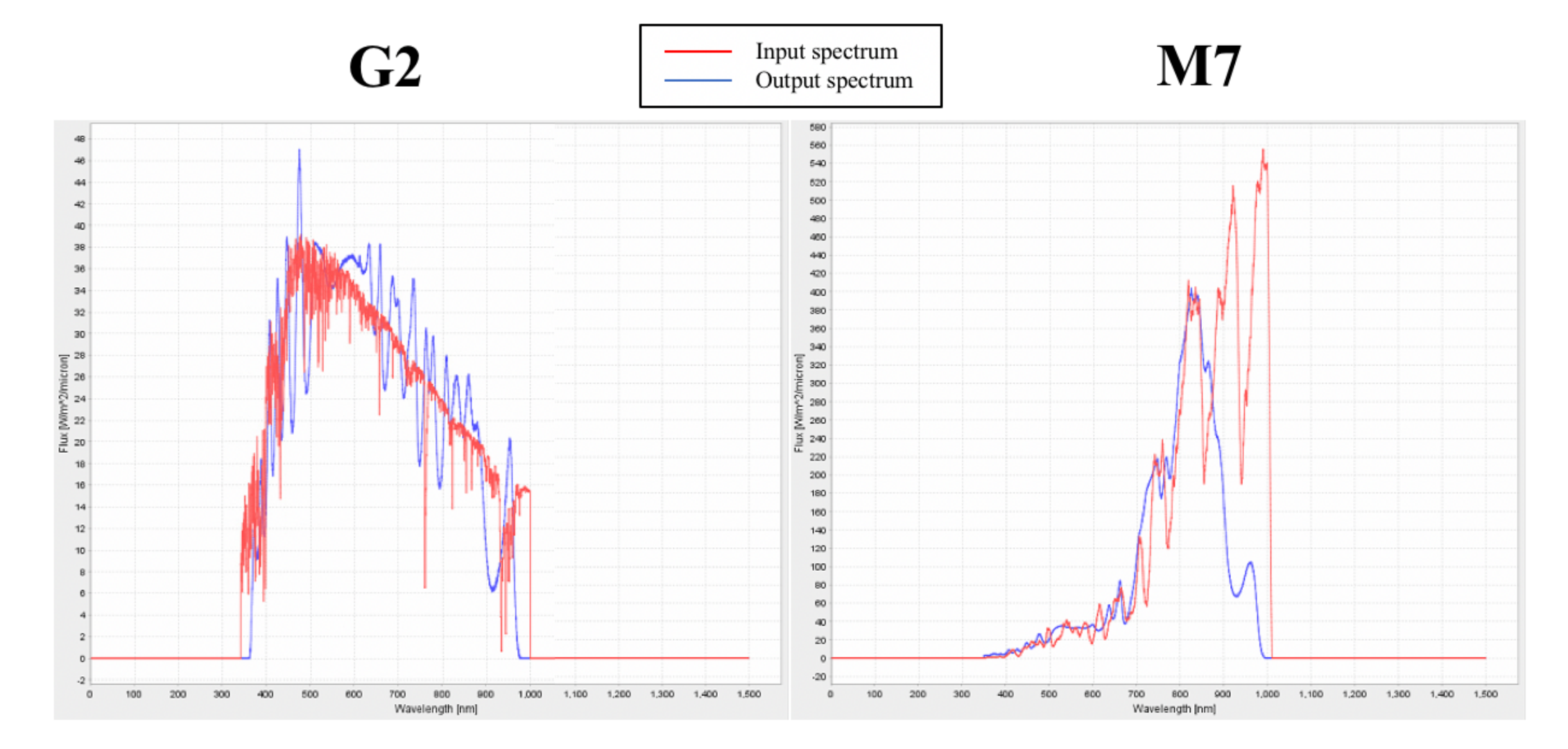}
\caption{\textbf{Left panel}: G2 V solar spectrum (input) in red and the emitted light spectrum (SOL%Please define if appropriate. RCl it is just a short for solar and it is the name of the light. It is not an acronym
) in blue. \textbf{Right panel}: M7 star spectrum (input) in red and the emitted light spectrum (M7) in blue. In~both panels, the spectra are plotted in W m$^{-2} \upmu$m$^{-1}$ versus wavelength (nm).}
\label{fig:guillu}
\end{figure} 

\begin{figure}[H]
\includegraphics[width=10 cm]{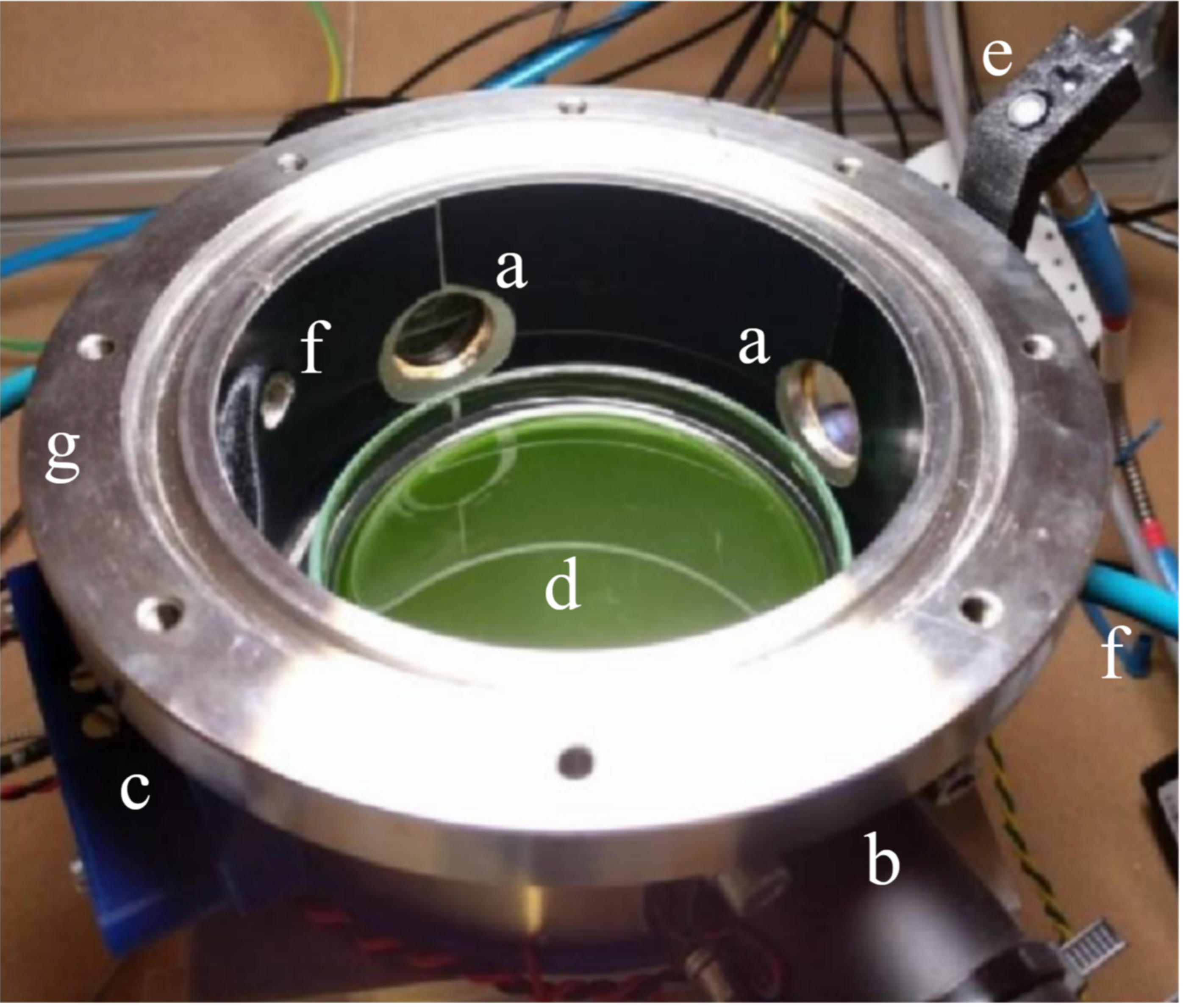}
\caption{Reaction cell inside the Starlight simulator box. Side windows (\textbf{a}) provide access to the CO$_2$ Tunable Diode Laser Absorption Spectroscopy (TDLAS) sensing channel (\textbf{b}) and to the fluorescence quenching O$_2$ sensor (\textbf{c}).~The reaction cell contains a Petri dish with target organisms inside (\textbf{d}). An optical fiber (\textbf{e}) provides feedback to the illuminator control spectrometer. The atmosphere inside the cell is flushed through input and output pipes (\textbf{f}). The flange (\textbf{g}) seals the cell with the top window (removed in this picture).}
\label{fig:recell}
\end{figure} 

In the context of this work, the cell was always operating at ambient pressure and 30~$^\circ$C temperature with an initial composition of 75\% vol. N$_2$, 20\% vol. O$_2$, and 5\% vol. CO$_2$; this~provides a high enough amount of carbon dioxide to be fixed into biomass throughout the experiment without excessively stressing the sample.%The atmosphere in the cell is set manually and could be changed by varying the levels of O$_2$, CO$_2$, and N$_2$. The cell is provided with pipe fittings and connected to an array of flow meters and needle valves (each for a different input gas: N$_2$, O$_2$, CO$_2$) to inject atmospheres of controlled and arbitrary compositions. It is kept at constant temperature by controlling a Peltier cell on which the cylinder is leaned. 
~Vital photosynthetic microorganisms in a liquid medium inside the cell are expected to produce oxygen. Hence, the cell is provided with a commercial fluorescence quenching oxygen sensor (Nomasense~O$_2$ P300), while the CO$_2$ concentration is monitored via a custom Wavelength Modulation Spectroscopy (WMS) Tunable Diode Laser Absorption Spectroscopy (TDLAS, \cite{danilovicetal2016spie}) set up. To~monitor the gas, four wedged windows of 2.5\ cm are pierced and paired two by two in opposite positions on the wall of the cell. Two of the windows are used by the CO$_2$ sensor for TDLAS, whereas one is used for the fluorescence quenching tablet, which is remotely sensed through an optical fiber.~The reaction cell underwent several modifications for reducing the systematic errors and the human interferences during the experiments. In~Battistuzzi~et~al.~(2020)~\cite{battistuzzietal2020fps}, a complete description of the very last version of the reaction cell and the results obtained from a biological point of view is presented.

\subsection{The Control Software}
The starlight simulator (illuminator and spectrometer) and the incubator cell environment (gas sensors and Peltier cell) are controlled by two separate processes, running on the same computer. We wanted a stand-alone software to control the simulator because it could also be used also for other laboratory applications (e.g., photo-bioreactors, microscopy, yeast growth \cite{erculianietal2015spie}).

\subsection{Validation of the Experimental Set Up}
To validate the experimental set up, we positioned the cyanobacterium \textit{Synechocystis} sp. Pasteur Culture Collection (PCC) 6803 liquid cultures into the simulator chamber with an atmosphere consisting of a mixture of gasses in the following composition: 75\% of N$_2$, 20\% of O$_2$, and 5\% of CO$_2$. Eventually, we irradiated it by means of the star simulator with a solar (G2 V) spectrum with three different intensities: 30, 45, and 95 $\upmu$mole\ m$^{-2}$\ s$^{-1}$ (from here, $\upmu$mole is used for $\upmu$mole of photons). The organisms exposed to different light regimes grew with good photosynthetic efficiency. 
%In this way, we were able to evaluate of the rate of the O$_2$ evolution and CO$_2$ fixation into biomass for the test organism. 
Description of the test and of the developed method to evaluate the growth of bacteria without any interferences by the operators are fully detailed in Battistuzzi et al. (2020) \cite{battistuzzietal2020fps}.

\section{Biological Experiment as Proof of Concept}
\label{sec:conc}

\subsection{Selected Organisms}
\label{sec:orga}
On Earth, the rise of photosynthetic organisms transformed in billions of years the primordial planet into our beloved and green planet. A review of the fundamental processes at play in photosynthetic organisms with a set of algae and plants on Earth is presented in Ref. \cite{kiangetal2007asbio}.
Among microbes, the most relevant in changing the characteristics of our planet were cyanobacteria that evolved oxygenic photosynthesis deeply influencing the Earth's atmosphere and leading to the ``oxygen-breathing life'' development. 
The recent finding of cyanobacteria able to use FR light for oxygenic photosynthesis due to the synthesis of chlorophylls $d$ and $f$, extending in vivo light absorption up to 750 nm \cite{ganandbryant2015envmi, Allakhverdievetal2015biochemistry}, suggests the possibility of exotic photosynthesis in planets around M stars.

We took advantage of the availability of a large selection of extremophiles, including cyanobacteria from soils, thermal spring muds and cave rocks, and cyanobacteria with UV-absorbing pigments. Extremophile cyanobacteria from environments characterized by low irradiances, rich in FR wavelengths \cite{ganandbryant2015envmi}, are selected for M dwarf star simulations. Some of these cyanobacteria are already known for coping with conditions not occurring in their natural environments, such as space and martian simulated conditions in low Earth orbit \cite{devera2014Ijasb, billietal2019asbio, billietal2013aim}. These astrobiological experiments pointed out that the limits of life have not been established well yet, and that extremophiles may have the potential to cope with the simulated environments planned in our experiment and, hence, can be a good source to identify potential astronomical biosignatures.

The selected extremophiles are expected to perform photosynthesis using the simulated M star light radiations, considering that the minimum light level for photosynthesis is about $0.01\ \upmu \text{mole}\text{ m}^{-2}\text{ s}^{-1}$, i.e., less than $10^{-5}$ of the direct solar flux at Earth in the PAR wavelength range (2000 $\upmu \text{mole}\text{ m}^{-2}\text{ s}^{-1}$ ) (Reference \cite{mckay2000grl, cockellandraven2004icarus} and References therein). This implies that, even at the orbit of Pluto, light levels exceed this value by a factor of 100. 

In particular, for our experiment, we chose two cyanobacteria species, \textit{Chlorogloeopsis~fritschii} PCC 6912 and \textit{Chroococcidiopsis thermalis} PCC 7203, which are able to perform FR light photoacclimation ((FaRLiP) \cite{ganandbryant2015envmi, wolfandblankenship2019photres}), and \textit{Synechococcus} PCC 7335, a peculiar cyanobacterium strain capable of both FaRLiP and Chromation Acclimation (CA). We~compared the responses of these species with those of \textit{Synechocystis} sp. PCC 6803, a~cyanobacterium unable to activate FaRLiP or CA, hence used as control organism.

\textit{Chlorogloeopsis\ fritschii} PCC 6912 is an organism that can thrive under various environmental conditions in terms of intensity and temperature. On Earth, its favorable habitats are thermal springs and hyper-salty lakes. The peculiarity of this strain is its ability to synthesize chlorophylls $a$, $d$, and $f$. Chlorophylls $d$ and $f$ are produced in a larger quantity when \textit{Chlorogloeopsis\ fritschii} grows in the FR light \cite{airsetal2014febsl}.

On the other hand, \textit{Chroococcidiopsis thermalis} PCC 7203 is a cyanobacterium isolated from a soil sample in Germany and ascribed to the species \textit{Chroococcidiopsis thermalis }Geitler, whose type locality is Sumatra hot springs (according to Algaebase website: \url{https://www.algaebase.org/}). Moreover, members of the \textit{Chroococcidiopsis} genus are widespread and can be found in freshwaters, salt waters, and hot and cold deserts \cite{cumbersandrothschild2014jp}. Due to their capability of withstanding different and extreme conditions, \textit{Chroococcidiopsis} strains are utilized in astrobiology studies \cite{billietal2019asbio}.
%that could be found in caverns, caves, hot and cold deserts, and hot springs. 
As for \textit{Chlorogloeopsis\ fritschii}, \textit{Chroococcidiopsis thermalis} performs FaRLiP, and both grow continuously and photoautotrophically in FR light and are utilized as model organisms to study cyanobacteria acclimation mechanisms to this light source \cite{zhaoetal2015fimcb}.

\textit{Synechococcus} sp. PCC 7335 was originally isolated from a snail shell in an intertidal zone and thus adapted to changes in light regimes and hydration/dehydration. This~organism is unique due to its capability to activate FaRLiP response when grown under FR light and to clearly show changing its pigmentations complementarily to the incident light spectrum, due to the so-called CA acclimation response \cite{hoetal2017phores}.

Regarding \textit{Synechocystis} sp. PCC 6803, it is a well-known cyanobacterium used as a model strain due to the complete sequencing of its genome. \textit{Synechocystis} sp. PCC 6803 has been selected as a control organism. In fact, it does not possess the gene cluster that is responsible for the FaRLiP, and it does not acclimate when exposed to FR light.

\subsection{Growth and Photosynthetic Efficiency}
\label{sec:grow}
To perform the first part of the experiment (see Section\ \ref{sec:exptim}), the selected cyanobacteria were grown in BG%Please define if appropriate.
-11 medium \cite{rippkaetal1979microbiology} or in ASN-III %Please define if appropriate.
medium \cite{hoetal2017phores}, depending on the species, in both liquid and solid (with the addition of 10 g l$^{-1}$ of Agar) cultures.~The~liquid cultures were exposed to terrestrial atmospheric air in a climatic chamber maintained at a temperature between 28 and 30 $^\circ$C under a continuous cool white fluorescent light of \mbox{$30\ \upmu$mole m$^{-2}$ s$^{-1}$} (L36W-840, OSRAM%Please define if appropriate. RCl: it is the name of the lamp factory
). Once the organisms were in the exponential growth phase, we subdivided them 
 into spots over agarized solid medium in petri plates with BG-11 or ASN-III, to be exposed to the different light sources: solar (SOL; see~\mbox{Figure~\ref{fig:guillu}}, left panel) with $20.3\ \upmu$mole m$^{-2}$ s$^{-1}$ in the PAR and $27.4\ \upmu$mole m$^{-2}$ s$^{-1}$ in the whole working range (380--780 nm), M7 V star (M7 see Figure~\ref{fig:guillu}, right panel) with \mbox{$20.3\ \upmu$mole m$^{-2}$ s$^{-1}$} in the PAR and $57.6\ \upmu$mole m$^{-2}$ s$^{-1}$ in the whole working range, and~a monochromatic flux centered at 720 nm (far-red (FR); see~Figure~S2 in the Supplementary Materials) with \mbox{$2.3\ \upmu$mole m$^{-2}$ s$^{-1}$} in the PAR and $20\ \upmu$mole m$^{-2}$ s$^{-1}$ in its working range (660--780~nm) (This source of light is a small lamp built with five red LEDs SMD (Surface Mount Device) OSLON 720.). The PAR luminosity for the simulated M star is close to the values evaluated by Reference \cite{kiangetal2007asbio}, incorporated in their Table 2. 

\begin{specialtable}[H] 
\caption{Values of F$_V$/F$_m$ obtained for several organisms. The considered error is $1 \sigma$.\label{tab:fvfm}}
\scalebox{1.13}[1.13]{\begin{tabular}{ccccc}
\toprule
\multirow{2}{*}{\textbf{Light Source}}			& \multicolumn{4}{c}{\textbf{PCC}} \\\cmidrule{2-5}
                                          &    \textbf{6803}           &  \textbf{6912 }         & \textbf{7203} &\textbf{7335}               \\
\midrule
SOL                                   &$0.568 \pm 0.050$&$0.513 \pm 0.018$&$0.501 \pm0.021$& $0.513 \pm 0.022$\\   
M7                                     &$0.578 \pm 0.008 $&$ 0.501 \pm 0.011$&$0.490 \pm 0.026$&$ 0.503 \pm 0.026$\\
FR                                     &$0.380 \pm 0.050 $&$ 0.523 \pm 0.028 $&$ 0.505 \pm 0.019 $&$ 0.560 \pm 0.013$\\
\bottomrule
\end{tabular}}
\end{specialtable}

For the experiment, we arranged plates with several spots of $20\ \upmu$L at different cell concentrations with 0.3, 0.5, 0.7, and 1 of optical density (OD) measured at 750 nm for each organism (see Figure\ S3 in the Supplementary Materials).~The plates were illuminated by the three different sources for 240 h and analyzed after 72 h and at the end of the experiment (see Figure\ S4 in the Supplementary Materials).~All the experiments were performed in three biological replicates.

The top panel of Figure\ \ref{fig:spot} shows the situation of the different phenotypes before and after 72 h and 240 h of irradiation with the three different light sources. An enhancement of the optical density of the spots of all the species that are greener than those at the beginning of the experiment can be observed in Figure\ \ref{fig:spot}. This trend is visible for both the sample irradiated by the solar light and the sample irradiated by the light with an M star simulated spectrum. 
The bacteria behaves differently from when they are exposed to the irradiation of the FR light as expected. In fact, it is possible to see that, unlike the other strains, the~control organism (PCC 6803) does not grow under the monochromatic FR light. 

\begin{figure}[H]
\includegraphics[width=10 cm]{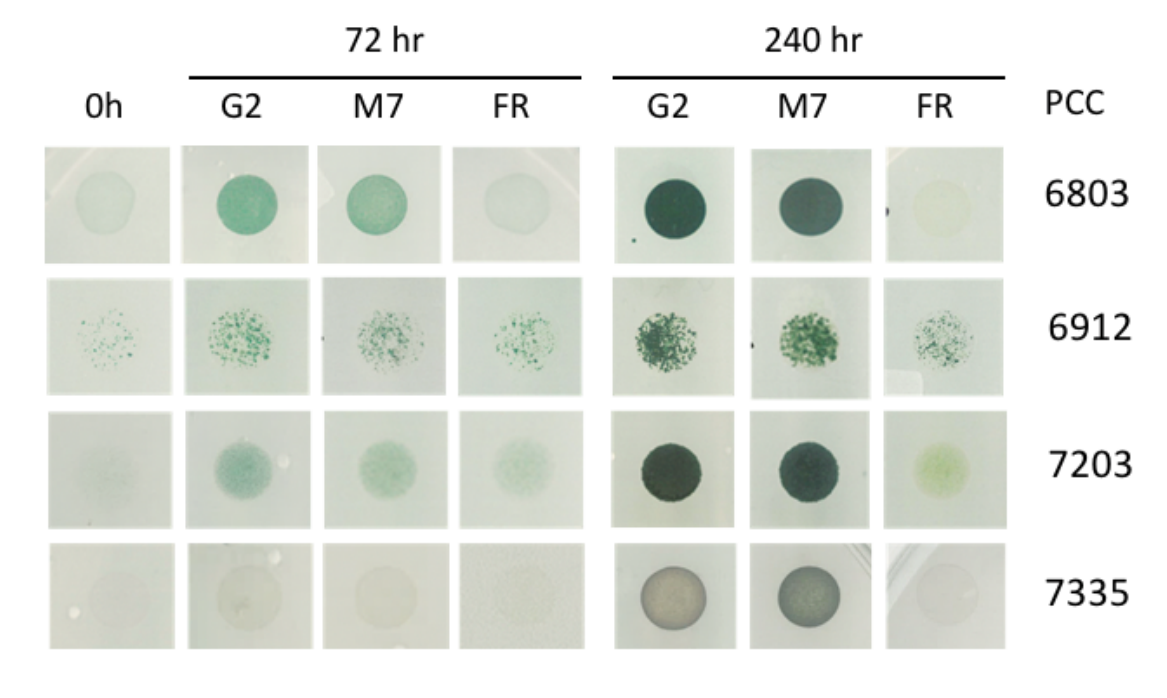}
\includegraphics[width=10 cm]{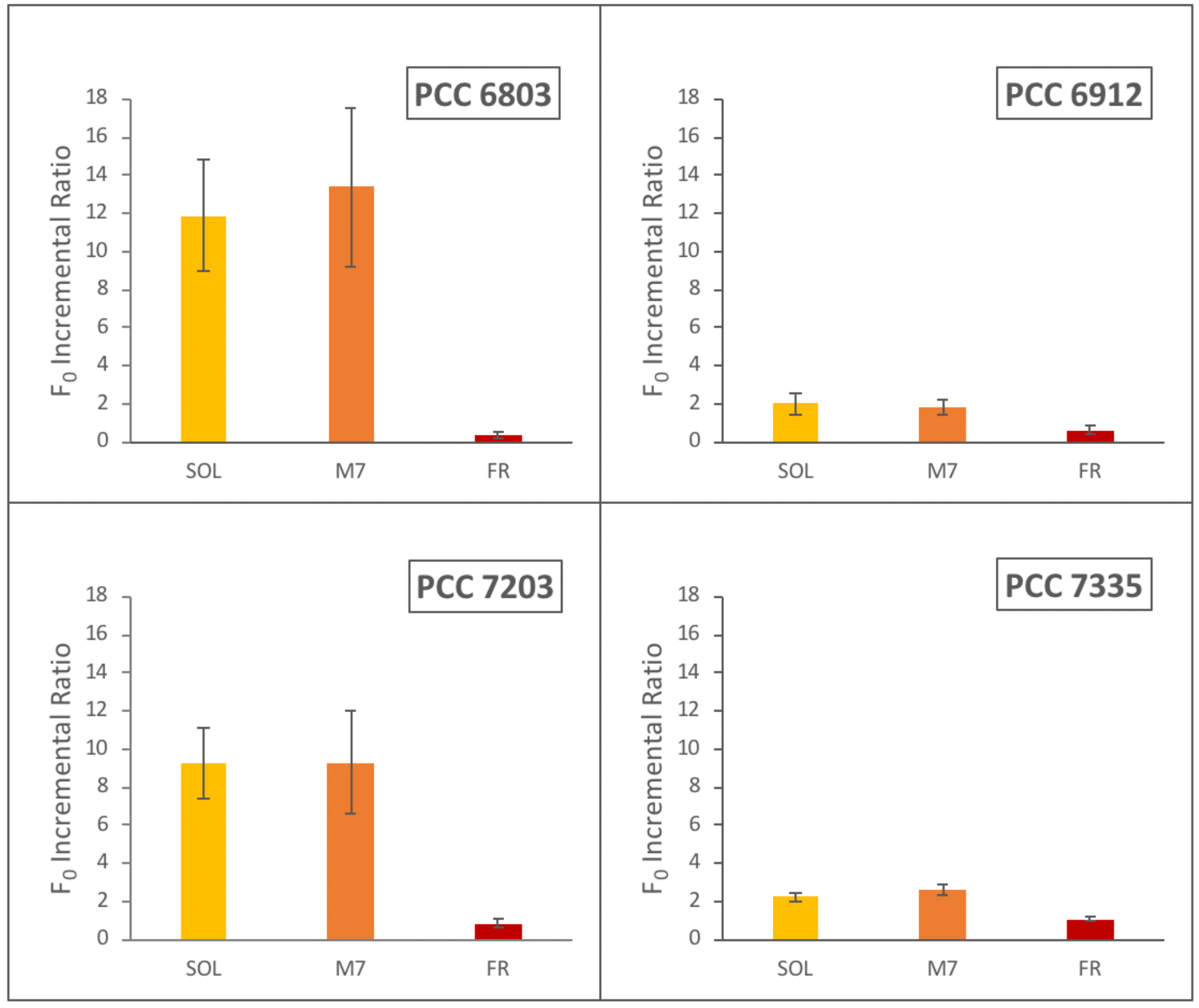}
\caption{\textbf{Top panel}: Photos of the strains in individual plates show the phenotypes of the organisms. From left to right, at the beginning (0 h), after 72 h, and 240 h of exposure to sunlight (SOL), a star M7 (M7), and far-red (FR). The phenotype variation of the spotted strains is related to those with 0.7 initial optical density (OD). \textbf{Bottom panel}: Values of the F$_0$ incremental ratio of the different organisms. The F$_0$ incremental ratio is defined as follows: ($\text{F}_{0(72H)}-\text{F}_{0(0H)}$)/$\text{F}_{0(0H)}$, with F$_{0(0H)}$ implying F$_0$ value at 0 h and
F$_{0(72H)}$ corresponding to F$_0$ value at 72 h. The F$_0$ incremental ratio values are reported in Table~S1 in the Supplementary Materials.}
\label{fig:spot}
\end{figure}

These results are confirmed and quantified by the values of the F$_0$ incremental ratios (Figure\ \ref{fig:spot}, bottom panel). F$_0$ is the ground chlorophyll fluorescence measured with Pulse-Amplitude Modulation (PAM) on cells adapted to the dark for 20--30 min. Eventually, they~are exposed to a pulse modulated light that does not trigger the photosynthetic process \cite{misumietal2016plantandcell}. The chlorophyll ground fluorescence (F$_0$) is proportional to the increment of chlorophyll, and thus, to the number of cells in the considered spot. 
So, an increase of the F$_0$ parameter is a measure of the growth of the culture \cite{perinetal2015biotech}.~The measurements are considered reliable when the F$_0$ value increases linearly with the increasing OD. For~this reason, we performed experiments with 4 different initial culture concentrations (OD) for each organism, and we repeated it with 3 independent biological replicates. The~best initial OD meeting these conditions proved to be 0.5, 0.7, and 1 for each organism.~The~measurements have been performed at the beginning of the experiment F$_{0(0H)}$ and after 72~h (F$_{0(72H)}$).~We~did not consider the measurement at 240 h as the signals of fluorescence with the initial PAM settings were saturated for most of the spots due to the very high cell concentrations.
F$_0$~chlorophyll fluorescence was measured on the entire spot at 0 and 72 h, by analyzing the same area and maintaining the same acquisition setting each time \cite{perinetal2015biotech}.
The F$_0$ detected at the beginning (F$_{0(0H)}$) and after 72 h (F$_{0(72H)}$) were used to calculate the F$_0$ incremental ratio ($\text{F}_{0(72H)}-\text{F}_{0(0H)}$)/$\text{F}_{0(0H)}$) and to estimate the growth.

The results reported in Figure\ \ref{fig:spot} (bottom panel) are concluded from spots with 0.7 initial OD and show the increase of F$_0$ after 72 h of different light sources exposure. The~histograms indicate that all cyanobacteria strains are capable of growing well under M7 simulated light, with incremental ratio F$_0$ comparable to those exposed to the solar light. PCC6803, that does not posses the FarLiP gene cluster, has an F$_0$ incremental ratio close to zero and negligible with respect to the values registered for M7 and solar light. All the other organisms are shown to be capable of exploiting FR light with different extents depending on the strain.

For evaluating the efficiency of the photosynthesis, we analyzed each plate by means of the PAM imaging (FluorCam FC 800MF). We evaluated the PSII %Please define if appropriate.
photosynthetic efficiency through the chlorophyll fluorescence measurements, before exposing the culture to the various light sources and after 72 h. The photosynthetic efficiency is defined by the ratio $\text{F}_V/\text{F}_m$, where $\text{F}_V=\text{F}_m-\text{F}_0$ with F$_0$ the basal fluorescence, and F$_m$ is the maximum fluorescence assessed after 20 min of dark adaptation followed by a flash of saturating light.
The averaged values of the parameters F$_ V$ /F$_ m$ obtained by the PAM analysis from three independent biological replicates for each organism, derived from the spots with 0.5, 0.7, and 1 initial OD are reported in Table \ \ref{tab:fvfm}. These values show that all the organisms maintain similar photosynthetic efficiency under different light regimes (with values typical of the cyanobacteria strains (Reference \cite{misumietal2016plantandcell, llewellyetal2020frontiers} in Supplementary Materials), while PCC6803 shows a decline of the F$_V$/F$_m$ parameter under FR light, as expected. 

\section{Discussion and Conclusions}
\label{sec:disconcl}
M stars are very popular in the astrobiology community due to their ubiquitous presence in the Galaxy and their small radii, which provide higher amplitude transit signals than solar-like stars. So far, they are recognized to be the most frequent hosts of super-Earths discovered orbiting in the HZ of a star. This sparked off a great theoretical debate about the possibility of having life, particularly photosynthetic life, on these planets. 
{Several~efforts have been spent aiming at modeling the upper wavelength limit of putative photoautotrophs on exoplanets. It has been hypothesized that oxygenic photosynthetic organisms could have developed pigments that do not utilize PAR light, but the more abundant NIR light, or employ photosystems using up to 3 or 4 photons per carbon fixed (instead of 2), as well as utilize more photosystems in series (3, 4, or even 6), allowing them to exploit photons of wavelengths up to 2100 nm \cite{wolstencroftandraven2002icar, kiang2007asbioi, kiangetal2007asbio, takizawa2017Rededge}. Moreover, the prospects for photosynthesis on habitable exomoons via reflected light from the giant planets that they orbit have been theoretically evaluated, suggesting that such photosynthetic biospheres are potentially sustainable on these moons except those around late-type M-dwarfs \cite{lingamandloeb2020ijasb}.}
However, up to now, no experimental data (survival, growth, and photosynthetic activity) about the behavior of oxygenic photosynthetic organisms exposed to simulated environmental conditions of exoplanets orbiting the HZ of M dwarfs, in particular, exposed to an M-dwarf light spectrum, have been produced. Numerous investigations (see Reference \cite{wolfandblankenship2019photres}, for~a review) have been done, instead, in the field of the oxygenic photosynthesis ``beyond 700 nm'', especially after the discoveries of cyanobacteria able to synthesize chl $d$ and $f$ \cite{miyashitaetal1996natur, chenetal2010sci}. However, they were committed to understanding the molecular and biochemical mechanisms behind how the photosynthetic functioning of the photosynthetic apparatus under a FR light spectrum, rather than testing it under exotic light spectra \cite{ganetal2014sci, zhaoetal2015fimcb, hoetal2017phores, hoetal2020bioch, nurnberg2018sci, hoandbryant2019, kurasovetal2019photres}.
Here, to the best of the authors' knowledge, for the first time, we present the experimental data obtained directly through exposing photosynthetic organisms to a simulated M dwarf spectrum. We compared the results to responses of those species under solar and FR simulated lights, using innovative laboratory instrumentation. As expected, in FR light, only the cyanobacteria able to synthesize chlorophyll $d$ and $f$ could grow. Surprisingly, all strains, both able or unable to use FR light, grew and photosynthesized under the M dwarf generated spectrum in a similar way to the solar light and much more efficiently than under the FR one.

In particular, we compared the responses of strains able to have FarLiP and of the control microorganism PCC 6803 that does not. The growth estimated by F$_0$ incremental ratio parameter obtained for all the cyanobacteria in our study shows a value that is very similar or equal, considering the error bars, to the value of F$_0$ measured for those spots irradiated with the solar light. In the case of the irradiation with monochromatic light (far-red (FR)), only PCC 6803 is unable to acclimate itself to the FR light, while all the others show a normal photosynthetic efficiency under this light, as well.~This suggests that PCC6803 grows very well under simulated M7 light by only using the visible part of the spectrum. The~ability of the other organisms to exploit FR light does not seem be beneficial for growth under M7 simulated light.
Furthermore, all the tested strains, except PCC 6803, have very similar values of the F$_V$/F$_m$ under any kind of irradiation spectra. This~highlights that they are able to acclimate to all the used lights. Our findings emphasize the importance of simulating both the visible and FR light components of an M dwarf spectrum to correctly evaluate the photosynthetic performances of oxygenic organisms exposed to such an exotic light condition.

Moreover, in a previous work \cite{battistuzzietal2020fps}, we demonstrated that, with our experimental set up, we can measure the consumption of CO$_2$ and the production of O$_2$ of the PCC 6803 cyanobacterium under solar irradiation. This serves as a prelude to the future analysis of the cyanobacteria photosynthetic gas exchanges in real time during their growth under M star spectra irradiation.

Last but not least, we realized an experimental set up that allows us to reproduce, in the laboratory, an alien environment with the possibility to variate the thermal and physical conditions. In this way, we are carrying out experiments on photosynthetic organisms to verify their capacity of thriving and acclimating to extraterrestrial conditions. We developed new and original laboratory devices (e.g., the star irradiation simulator) and novel measurement methods (see Reference \cite{battistuzzietal2020fps}) that will allow new experiments in the future. To prepare for the next step of our research plan, we have already produced several models of stable super-Earth atmospheres to be used in the laboratory. We have started to monitor the evolution of oxygen and the fixation of carbon dioxide in the cyanobacteria exposed to very different irradiations and simulated atmospheres. 

Hence, if the evolutionary tracks on a habitable planet in the HZ of an M star are quite similar to those on Earth, photosynthetic microorganisms could, as well, produce O$_2$ and fix CO$_2$ in organic matter on these planets orbiting such cold stars. %The outcome of our experiments is that it does not matter if the photon density in the PAR range is as low as it can be low for M stars, the cyanobacteria can use all these photons in an efficient way without any kind of stress.

Will it be possible to observe the released oxygen in a remote way? The answer to this question is not simple because it depends not only on the efficiency in producing oxygen by photosynthetic organisms but also on the efficiency of the possible oxygen sinks that are at work on that planet. The reverse reaction to oxidize photosynthetic products depletes the atmospheric oxygen. The net release of oxygen in the atmosphere, due to this balance, is regulated by the sink of organics in the sediments. If the level of O$_2$ is low in the atmosphere, the reactions with reducing gases from vulcanism (H$_2$~and H$_2$S) and submarine \mbox{weathering \cite{catlingandkasting2017aeil.book, kalteneggeretal2010asbio}} can deplete O$_2$. If the O$_2$ production rate is greater than the depletion rate, its build-up in the atmosphere is possible \cite{lehmeretal2018apj}, and the Fe$^{2+}$ oxidation process becomes an important one. Catling and Kasting (2017) and Kaltenegger~et~al.~(2010)~(Reference \cite{catlingandkasting2017aeil.book, kalteneggeretal2010asbio}, respectively, and~References therein) discussed deeper on the build-up of oxygen in the atmosphere of a planet. Oxygen depletion is a time-dependent process. The atmospheric oxygen is recycled through respiration and photosynthesis in less than 10,000 years. In the case of total extinction of the biosphere of Earth, the atmospheric O$_2$ would disappear in a few million years \cite{kalteneggeretal2010asbio}.

Thus, we conclude that only the observations can give us the right answer. So far, brand new ground- and space-based instruments are planned to be operative with the aim of finding and characterizing extrasolar planets. In the next future, dedicated space missions and space telescopes, like James Webb Space Telescope (JWST) and Origin Space Telescope (OST), and huge ground telescopes will be the right tools to search for life in other worlds.

%%%%%%%%%%%%%%%%%%%%%%%%%%%%%%%%%%%%%%%%%%

%%%%%%%%%%%%%%%%%%%%%%%%%%%%%%%%%%%%%%%%%%
%\section{Patents}
%This section is not mandatory, but may be added if there are patents resulting from the work reported in this manuscript.

%%%%%%%%%%%%%%%%%%%%%%%%%%%%%%%%%%%%%%%%%%
\vspace{6pt} 

%%%%%%%%%%%%%%%%%%%%%%%%%%%%%%%%%%%%%%%%%%
%% optional
\supplementary{The following  are available online at https://www.mdpi.com/2075-172 9/11/1/10/.
Figure S1: M7 V star input spectrum as it appears before (in light gray), and after (in red) the smoothing process. The emitted spectrum (in blue) is superimposed. The smoothing process reduces the resolution of the input spectra. Hence, the stellar simulator is able to better reproduce the input spectrum following the depletion of flux due to large atomic absorptions or molecular bands. Figure S2: Emitted FR light measured by the FLAME VIS-NIR spectrograph. The central wavelength is 720\ nm and the full width at zero level is 130\ nm with a wavelength range 650\ nm and 780 nm. The luminosity of this lamp is $2.3\ \upmu$mole m$^{-2}$ s$^{-1}$ in the PAR and $20\ \upmu$mole m$^{-2}$ s$^{-1}$ in the total working range. Figure S3: Different cultures of the selected cyanobacteria with different optical density, before the 20$\mu$l spots were deposited on the Petri's plate. Figure S4: Examples of a BG-11 agar plate with S. sp. PCC6803, \textit{C. fritschii} PCC6912 and \textit{C. thermalis} PCC 7203 spots and a BG-11-ASN III agar plate with S. sp. PCC7335. Plates are shown after 72 and 240 h of exposure under the M-dwarf simulated spectrum. Table S1: Averaged values (n$=6$) of  the F$_0$ incremental ratio obtained for several organisms under different light sources. The considered error is $1 \sigma$. }
\authorcontributions{R.C., N.L.R., and L.P. made the conceptualization and supervision of the whole experiment. R.C., M.E., L.C., M.B., N.L.R., and L.P. define the methodology necessary to perform the experiment. E.A. and R.C. realized the super-Earths atmospheric models. R.C., L.P., L.C., and M.E. conceived and realized the Star Light Simulator (SLS) and the Atmosphere Simulator Chamber (ASC). B.S. realized the control software of the S.L.S. and L.C. the control software of the ASC. B.S., M.E., and E.A. calibrated the S.L.S., M.B., N.L.R., A.C.P., and D.S. conceived the biological experiments. A.C.P., D.S., and M.B. performed them and analyzed the data. L.C., R.C., M.E., L.P., and N.L.R. contributed to the analyses. R.C. wrote the manuscript, NLR edited it, and E.A., M.B., L.C., M.E., A.C.P., B.S., V.S., L.P. critically read it. All authors have read and agreed to the published version of the manuscript.}

%%%%%%%%%%%%%%%%%%%%%%%%%%%%%%%%%%%%%%%%%%
\funding{The research was co-funded by the Italian Space Agency through the ''Life in Space'' project (ASI N. 2019-3-U.0); by the Progetti Premiali scheme (Premiale WOW) of the Italian national Ministry of Education, University, and Research; by the Department of Biology of University of Padova and the Institute for Photonics and Nanotechnologies of CNR through intramural grants.}

\institutionalreview{Not~applicable}

\informedconsent{Not applicable}

\dataavailability{Data is contained within this article and supplementary material} 

%%%%%%%%%%%%%%%%%%%%%%%%%%%%%%%%%%%%%%%%%%
\acknowledgments{The authors would like to thank T. Morosinotto and G. Galletta for their very useful comments and suggestions, as well as F. Z. Majidi for her fundamental help in editing the final version of the paper. The authors would like to thank also the two referees for their useful comments.}

%%%%%%%%%%%%%%%%%%%%%%%%%%%%%%%%%%%%%%%%%%
\conflictsofinterest{The authors declare no conflict of interest. The funders had no role in the design of the study; in the collection, analyses, or interpretation of data; in the writing of the manuscript, or in the decision to publish the results.} 

\newpage
%%%%%%%%%%%%%%%%%%%%%%%%%%%%%%%%%%%%%%%%%%
%% optional
\abbreviations{The following abbreviations are used in this manuscript:\\

\noindent 
\begin{tabular}{@{}ll}
CA & Chromatic Acclimation\\
CCT & Correlated Color Temperature\\
CHL & Chlorophyll \\
COTS & Component Off The Shelf\\
FaRLiP & Far-red light photoacclimation \\
HZ & Habitable Zone\\
JWST & James Webb Space Telescope\\
NIR & near infra-red \\
OD & Optical Density\\
OST & Origin Space Telescope\\
PAM & Pulse-Amplitude Modulation\\
PAR & Photochemically Active Radiation\\
PC & Personal Computer\\
PCC & Pasteur Culture Collection\\
TDLAS & Tunable Diode Laser Absorption Spectroscopy\\
WMS & Wavelength Modulation Spectroscopy\\
\end{tabular}}

%%%%%%%%%%%%%%%%%%%%%%%%%%%%%%%%%%%%%%%%%%
%% optional
%\appendixtitles{no} % Leave argument "no" if all appendix headings stay EMPTY (then no dot is printed after "Appendix A"). If the appendix sections contain a heading then change the argument to "yes".
%\appendix
%\section{}
%\unskip
%\subsection{}
%The appendix is an optional section that can contain details and data supplemental to the main text. For example, explanations of experimental details that would disrupt the flow of the main text, but nonetheless remain crucial to understanding and reproducing the research shown; figures of replicates for experiments of which representative data is shown in the main text can be added here if brief, or as Supplementary data. Mathematical proofs of results not central to the paper can be added as an appendix.

%\section{}
%All appendix sections must be cited in the main text. In the appendixes, Figures, Tables, etc. should be labeled starting with `A', e.g., Figure A1, Figure A2, etc. 

%%%%%%%%%%%%%%%%%%%%%%%%%%%%%%%%%%%%%%%%%%
\end{paracol}
\reftitle{References}

\end{document}